%% file: main_R1.tex
\documentclass[fleqn,usenatbib]{mnras}

\usepackage{newtxtext,newtxmath}

\usepackage[T1]{fontenc}

\DeclareRobustCommand{\VAN}[3]{#2}
\let\VANthebibliography\thebibliography
\def\thebibliography{\DeclareRobustCommand{\VAN}[3]{##3}\VANthebibliography}

\usepackage{graphicx}	
\usepackage{amsmath}	
\usepackage{multirow}

\newcommand{\teff}{$\mathrm{T_{eff}}$}
\newcommand{\logg}{$\log$ g}
\newcommand{\vsini}{$v\sin i$}

\newcommand{\tona}{{\tt TONALLI}}
\newcommand{\alfa}{[$\alpha$/Fe]}
\newcommand{\meta}{[M/H]}
\newcommand{\bacc}{{\tt BACCHUS}}
 
\title[Parameters and abundances of YSOs I. Orion]{Atmospheric parameters and chemical abundances of young stars with APOGEE. I. Orion star-forming region}

\author[L\'opez-Valdivia et al.]{Ricardo L\'opez-Valdivia,$^{1}$\thanks{E-mail: rlopezv@astro.unam.mx}
Luc\'ia Adame,$^{1}$
Carlos G. Rom\'an-Z\'u\~niga,$^{1}$
Jes\'us Hern\'andez,$^{1}$
Edilberto S\'anchez,$^{1}$
\newauthor{Itzarel Herrn\'andez-Aburto,$^{1}$}
Jos\'e G. Fern\'andez-Trincado,$^{2}$
Eduardo Zagala Lagunas,$^{3}$
Leticia Carigi,$^{4}$
\newauthor{J. E. M\'endez-Delgado,$^{4}$}
Marina Kounkel,$^{5}$
Javier Serna,$^{6}$
Richard R. Lane,$^{7}$
Keivan G. Stassun,$^{8}$
Sandro Villanova,$^{9}$
\newauthor{Jinyoung Serena Kim,$^{10}$}
{S. J. Wolk,$^{11}$}
Guy S. Stringfellow,$^{12}$
Jonathan C. Tan,$^{13,14}$
A. Roman-Lopes,$^{15}$
\newauthor{B\'arbara Rojas-Ayala,$^{16}$}
Rakesh Pandey,$^{17}$
\\
$^{1}$Instituto de Astronom\'ia, Universidad Nacional Aut\'onoma de M\'exico, Ap. 106,  Ensenada 22800, BC, M\'exico\\
$^{2}$Universidad Cat\'olica del Norte, Instituto de Astronom\'ia, Av. Angamos
0610, Antofagasta, Chile \\
$^{3}$Facultad de Ciencias, Universidad Aut\'onoma de Baja California, 22860 Ensenada, B. C., M\'exico\\
$^{4}$Instituto de Astronom\'ia, Universidad Nacional Aut\'onoma de M\'exico, Ap. 70-264, 04510 CDMX, M\'exico \\
$^5$Department of Physics and Astronomy, University of North Florida, 1 UNF Dr, Jacksonville, FL 32224, USA \\
$^6$Homer L. Dodge Department of Physics and Astronomy, University of Oklahoma, Norman, OK 73019, USA\\
$^7$Centro de Investigaci\'on en Astronom\'ia, Facultad de Ingenier\'ia, Ciencia y Tecnolog\'ia, Universidad Bernardo O'Higgins, Avenida Viel 1497, Santiago, Chile\\
$^8$Department of Physics and Astronomy, Vanderbilt University, VU Station 1807, Nashville, TN 37235, USA\\
$^9$Universidad Andres Bello, Facultad de Ciencias Exactas, Departamento de F{\'i}sica y Astronom{\'i}a - Instituto de Astrof{\'i}sica, Autopista
Concepci\'on-Talcahuano 7100, Talcahuano, Chile \\
$^{10}$Steward Observatory, University of Arizona, 933 N. Cherry Ave., Tucson,
AZ 85721-0065, USA \\ 
$^{11}$Center for Astrophysics | Harvard \& Smithsonian, 
60 Garden Street, Cambridge, MA 02138 USA \\
$^{12}$Center for Astrophysics and Space Astronomy, University of Colorado Boulder, Boulder, CO 80309, USA \\
$^{13}$Dept. of Astronomy, Univ. of Virginia, Charlottesville, VA 2904, USA\\
$^{14}$Dept. of Space, Earth \& Environment, Chalmers Univ. of Technology, E-412 96 Gothenburg, Sweden\\
$^{15}$ Departamento de Astronomía, Facultad de Ciencias - Universidad de La Serena, Av. Raul Bitran 1302, La Serena, Chile\\
$^{16}$Instituto de Alta Investigaci\'on, Universidad de Tarapac\'a, Casilla 7D, Arica, Chile \\
$^{17}$ Universidad Nacional Aut\'onoma de M\'exico, Instituto de Radioastronom\'ia y Astrof\'isica, Antigua Carretera a P\'atzcuaro 8701, Ex-Hda. \\
San Jos\'e de la Huerta, 58089 Morelia, Michoac\'an, M\'exico\\
}
\date{Accepted XXX. Received YYY; in original form ZZZ}

\pubyear{2015}

\begin{document}
\label{firstpage}
\pagerange{\pageref{firstpage}--\pageref{lastpage}}
\maketitle

\begin{abstract}
We derive atmospheric parameters and chemical abundances in young G-, K-, and M-type 
stars (temperatures between 6500 and 3100 K) using infrared APOGEE-2 spectra. 
Atmospheric parameters were determined for 548 young stars in the Orion complex (Orion 
A, B, OB1, and $\lambda$ Ori) using the \tona\ code. For 340 slow rotators (\vsini\ 
$\leq$ 30 km s$^{-1}$), we derived C, Mg, Si, K, Ti, and Fe abundances using 19 atomic 
lines, MARCS model atmospheres, and \bacc. To mitigate the impact of circumstellar 
material, we excluded stars with infrared excess identified via 2MASS and WISE 
photometry. We find sub-solar [X/H] abundance ratios, consistent across elements and 
among all four groups, suggesting a chemically homogeneous Orion complex. We computed 
[$\alpha$/Fe] from [Mg/Fe], [Si/Fe], and [Ti/Fe], obtaining a median of $-0.14 \pm 
0.04$, about 0.10 dex lower than the value for nearby main-sequence stars ($-0.04 \pm 
0.04$) at similar [Fe/H]. This result aligns with predictions from Galactic chemical 
evolution models. Furthermore, the median [C/H] abundance we derived for Orion agrees 
with previous estimations based on the analysis of the ionized gas of the Orion 
nebula. This work sets the stage for extending the analysis to stars with circumstellar 
material and higher rotational velocities, which will not only improve our 
understanding of Orion, but also provide critical insight into the formation and 
evolution of young stars, as well as the chemical evolution of the Milky Way. 
\end{abstract}

\begin{keywords}
stars: abundances -- stars: fundamental parameters -- stars: pre-main-sequence -- infrared: stars 
\end{keywords}


\section{Introduction}

In recent years, thanks to large-area spectroscopic surveys like the RAdial Velocity Experiment \cite[RAVE, ][]{rave03, rave20}, the Large Sky Area Multi-Object Fiber Spectroscopic Telescope \citep[LAMOST, ][]{cui12,zhao12}, the Gaia-ESO \citep{gilmore12}, the GALactic Archeology with Hermes \citep[GALAH, ][]{silva15,martell17}, and the Sloan Digital Sky Survey IV \citep[SDSS-IV;][]{blanton17}APOGEE-2 \citep{majewski17} and SDSS-V \citep{almeida23} Milky Way Mapper \citep{kollmeier17}, we have access to enormous inventories of stellar spectra with diverse stellar populations. These and other surveys are designed to provide uniform quality data, with sufficient signal-to-noise (SNR) and spectral resolution to allow us the extraction of reliable stellar parameters, like effective temperature (\teff), surface gravity (\logg), overall metallicity (\meta), age, mass, among others.

The core samples of most of the mentioned surveys were stars in the main sequence or post-main-sequence, with relatively quiescent photospheres that allow comparing properties of stellar populations in distinct Galactic components. However, pre-main sequence or very young stars (age $<$ 300 Myr) are subject to various processes that complicate the determination of their properties directly from their spectra, like high stellar rotation, magnetic activity, variability, among others.  

The SDSS-IV APOGEE-2 survey, which operates in the $H$-band ($\sim$1.5 to 1.7$\mu$m), provides a valuable window into the stellar photospheres of young stars. This wavelength range is significantly less affected by interstellar extinction and continuum veiling than the optical regime, making it especially suitable for studying recently formed stars in nearby star-forming regions. Consequently, APOGEE-2 data has become an increasingly used tool in the investigation of young stellar populations \citep[e.g., ][]{cottar14,foster15,cottar15,fernandez17,roman19,roman23}. 
However, a persistent challenge remains: {the reliable ($\sigma \lesssim 0.1$ dex) determination of chemical abundances in young stars. Although early studies \citep[e.g.,][]{padgett96,santos08} achieved abundance uncertainties slightly above 0.1 dex, they were limited by small sample sizes, which prevented more detailed chemical analyses.}

The Orion star-forming region, one of the most studied regions due to its youth and number of stars \citep[e.g., ][]{blaauw64,brown95,stassun99,dario16,dario17,kounkel18,hernandez23}, has been the focus of ongoing debate regarding its chemical homogeneity. The answer to this debate is of particular interest in better understanding and constraining models of the chemical evolution of star-forming regions.  The topic is still open. For example, \cite{cunha94} determined the abundance of C, N, O, Si, and Fe for 18 B-type main-sequence stars located in four groups of different ages. They found that the abundance of O and Si was higher, up to 40$\%$, in some of the youngest stars analyzed, but found no enrichment in C, N, or Fe. The authors suggested the observed trend to the enrichment of the molecular cloud (where such stars formed) with the ejecta material of type {\sc ii} supernovae, which is rich in O and Si but not in C or N. \cite{cunha98} found a similar trend in F and G Orion stars. In contrast, \cite{simon10} showed that the Si and O abundances of B stars in four subgroups of Orion OB1 were all consistent with solar values. Later, \cite{dorazi09} found that stars in the Orion OB1 group show a lower Fe abundance value than younger, more embedded regions like the Orion nebula cluster. Subsequent studies of \cite{biazo11, biazo12} confirmed, through the abundance analysis of several elements, that Orion is consistent with a thin-disk population. 

{More recently, \citet{kos21} revisited the chemical inhomogeneity in Orion using a significantly larger sample of GALAH spectra, consisting of about 300 stars across 15 different stellar clusters. Their results showed no evidence of self-enrichment in the $\lambda$-Ori group relative to the Orion B and Orion OB groups, suggesting that the Orion complex is chemically homogeneous.} It is important to note that \citet{kos21} reported strong correlations between elemental abundances and \teff, similar to those found by \citet{viana09} for Fe. These trends may indicate unaccounted factors such as stellar variability, departures from Local Thermodynamic Equilibrium (LTE), model dependencies, inaccurate atomic data, line blending, among others.

The present study aims to present the chemical abundances of young stars in four different groups of Orion through a homogeneous and consistent method. We take advantage of the quantity and quality of the near-infrared spectra of the SDSS-IV APOGEE-2 survey, and our goal is to contribute to the understanding of the chemical evolution of star-forming regions.

\section{Orion sample}
\cite{roman23} presented a catalog consisting of 3360 young stars, ranging from sub-solar to super-solar spectral types, located in 16 Galactic star-forming regions and young open clusters observed by the APOGEE-2 survey \citep{majewski17}. 
\cite{roman23} selected the young stars in this catalog using a clustering method that 
removed most field contamination, yielding a robust sample of bona fide members of 
their respective star-forming regions.

To define our working sample, we retrieved all epoch-combined 1D infrared spectra of the 1861 Orion members identified by \cite{roman23} from the APOGEE-2 DR17 public database.\footnote{We used the apStar spectra downloaded from the SDSS-IV science archive webapp.}
Young stars often exhibit infrared excesses caused by interstellar extinction or the presence of a protostellar disk. Both extinction and disk-related phenomena relate to additional physical processes, including accretion, magnetic activity, and continuum veiling, that can significantly affect the determination of atmospheric parameters and chemical abundances. For instance, continuum veiling is a non-stellar emission that reduces the depth of photospheric absorption lines in the spectra of young stars \citep{joy49}. In the infrared, veiling is attributed to excess emission from dust at the inner disk wall's sublimation radius \citep[e.g.,][]{natta01, muzerolle03} and to warm gas within that radius \citep[e.g.,][]{fischer11}.

Determining the veiling is a complex task that lies beyond the scope of this paper. However, we minimized its impact by identifying and excluding from our sample those stars exhibiting infrared excess, {following a semi-empirical approach.
We collected 2MASS and WISE photometry for a sample of main-sequence stars brighter than $J < 13$. The sample was divided into 15 bins of $J$-band magnitude, and a Gaussian function was fitted to the corresponding ($K - W3$) color distributions to determine their expected photospheric flux values. The results showed that, for bright stars ($J < 8$ mag), the photospheric flux remains approximately constant at 0.267 mag. For stars in the range $8 < J < 13$, the photospheric flux is better described by the following  fourth-degree polynomial.}

\begin{equation} 
    (K-W3)_{\rm phot} =  a + bJ+cJ^2+dJ^3+eJ^4
\label{eq:poli}
\end{equation}

where $a=0.5138, b=-0.2258, c=0.0713, d=-0.0091$, and $e=0.0004$. We looked for the 2MASS \citep{2mass} and 
WISE \citep{wise} photometry for the young stars of Orion in the {\tt ALLWISE} 
catalog \citep{cutri13}. The {\tt ALLWISE} catalog provides the WISE photometry quality 
and contamination flags, which are useful to identify bad WISE data. The quality flag (qph) measures the quality of the profile fit photometry, based on the signal-to-noise ratio of each band, while the contamination flag (ccf) indicates if the photometry may be affected by an image artifact. We found that 559 sources had reliable (qph equal to A, B, or C) and unaffected (ccf = 0) WISE photometry and ($K-W3$) consistent, within 1.5 times their respective error, with their corresponding photospheric value. These stars comprised our final working sample. 

\begin{figure}
    \centering
    \includegraphics[width=\columnwidth]{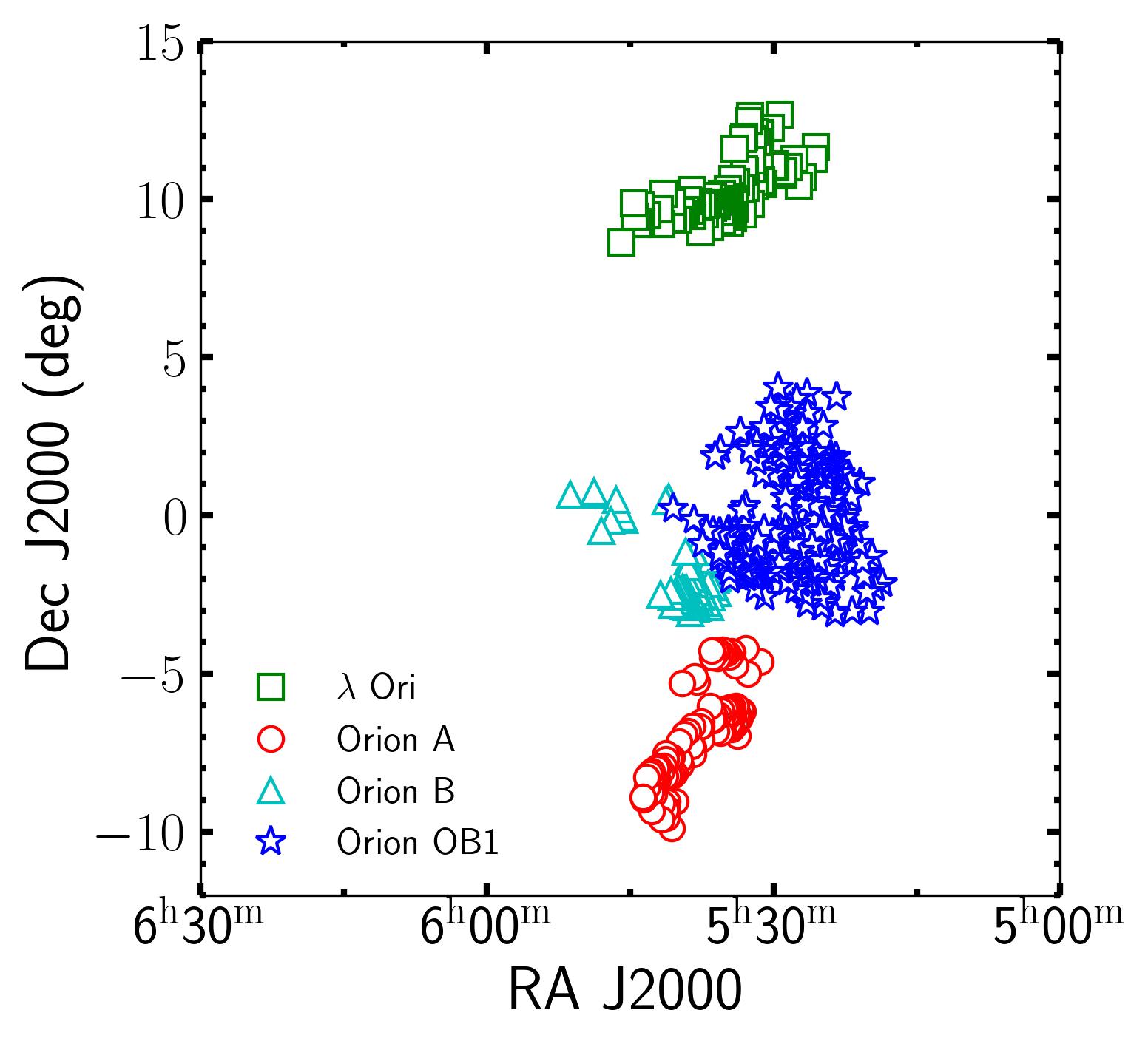}
    \caption{Spatial distribution of the 559 young stars located in the Orion star-forming region analyzed in this work.}
    \label{fig1}
\end{figure}

Most of our young stars are in the Orion OB1 group (228 stars). The remaining stars are located mostly uniformly in the $\lambda$ Ori (116), Orion A (112), and Orion B (103) groups. The spatial distribution of our young sample is presented in Figure~\ref{fig1}.

\section{Atmospheric parameter determination}
In this section, we summarize the methods followed to determine the atmospheric parameters of the Orion stars. We recommend the reader see a more detailed 
explanation about \tona\ in \cite{adame24} and on using \tona\ in a stellar sample of low-mass main-sequence stars in \cite{lopezval24}.

\subsection{\tona}
To determine the basic atmospheric parameters of our sample, we employed the code \tona\ \citep{adame24}. 

In short, \tona, which is based on the Asexual Genetic Algorithm of \citet*{canto09}, {randomly selects a set of stellar parameters and interpolates a synthetic spectrum at those values using the synthetic spectral grid of \citet{jonsson20}. The resulting synthetic spectrum is then rotationally broadened using the kernel described by \citet{gray08}, assuming a limb darkening coefficient of 0.4. Finally, it is compared with the observed spectrum using a $\chi^2$ statistic.} The {stellar parameters (including \vsini)} that produce the lower $\chi^2$ values are selected as ``fathers'' to create the next generation of synthetic spectra. In the next generation, the synthetic spectra are created by randomly selecting spectra from a small vicinity around each father, and these new synthetic spectra are compared again to the observed spectrum, finding new fathers. The search vicinity around the fathers decreases in size each generation until the stop criteria are achieved and the best-fit spectrum is found. 

The atmospheric parameters fitted by \tona\ are variable, and they are directly related to the theoretical grid used. In our case, we fitted the effective temperature (\teff), surface gravity (\logg), overall metallicity (\meta)\footnote{This value comprises the sum of all metals.}, $\alpha$-elements enhancement, and the projected rotational velocity (\vsini). The first four atmospheric parameters are bounded by the limits of the \cite{jonsson20} grid, while the allowed range for \vsini\ is set by the user.

{It is important to note that \citet{jonsson20} specifically developed a set of synthetic spectral grids tailored for APOGEE data. These grids are based on MARCS model atmospheres \citep{marcs}, match the average spectral resolution of APOGEE observations, and are synthesized at five different microturbulence velocities ($\xi$).
For our analysis, we used the grid computed at $\xi = 2.4$ km s$^{-1}$ and explored the full parameter space coverage for \teff, \meta, and \alfa\ (see Table~\ref{tab:space})}. For the \vsini, we constrained our fitting procedure to \vsini\ values between 0 and 50 km s$^{-1}$. Although pre-main-sequence stars might have rotational velocities higher than 50 km s$^{-1}$, their atmospheric parameters and chemical abundance determinations become more challenging due to the line blending produced in the spectrum by the rotation. 

Regarding \logg, we used {\tt TEPITZIN}, a photometric module incorporated in the \tona\ 
ecosystem, to obtain a photometric \logg\ value that could be used as a prior (Adame et al. in 
prep). {\sc tepitzin} minimizes the distances between the observed photometric colors (2MASS 
and Gaia) and the theoretical ones of the PARSEC pre-main-sequence evolutionary models 
\citep{parsec} to determine a best-fit photometric \logg\ value. This photometric \logg\ value serves as the center of a prior that is given to \tona. To keep consistent with the rest of 
the atmospheric parameters, we used a flat prior of 1 dex of total width, however, the user 
could choose a Gaussian or other functional form for this prior. 

Most of the young stars of Orion might still be, at least partially, embedded into their 
progenitor cloud producing a reddening that affects their photometry. Reddening  
would lead to inaccurate \logg\ photometric values. For this reason, it is necessary to 
correct the photometry of our sample before using it in {\sc tepitzin}. 

To determine the visual extinction ($A_v$) of our sample, we implemented the code {\sc massage} 
\citep{hernandez23}. {\sc massage} requires temperature, Gaia (Gp, Rp, and Bp), and 2MASS (J and
H) photometry to compute $A_v$ by minimizing the differences between the observed and expected 
intrinsic colors of \cite{parsec} affected by reddening. The extinction $A_v$ 
is changed until the best comparison is found through a $\chi^2$ statistic. We determined $A_v$ 
for our sample, de-reddened their photometry, and then used the extinction corrected 
photometry into {\sc tepitizin} to obtain a prior on \logg. 


\subsection{Atmospheric parameters}
Once we have the extinction corrected photometry, we determined the final atmospheric parameters 
(and their errors) of our sample performing 30 repetitions in \tona. We used the median (50 
percentile) of the posterior distributions as the value of the parameter analyzed. We took the
interquartile range of the results as the associated error\footnote{The interquartile range is 
defined as the difference between the 75 and 25 percentiles of the posterior distribution.} to
each parameter. 

\begin{figure*}
    \centering
    \includegraphics[width=\textwidth]{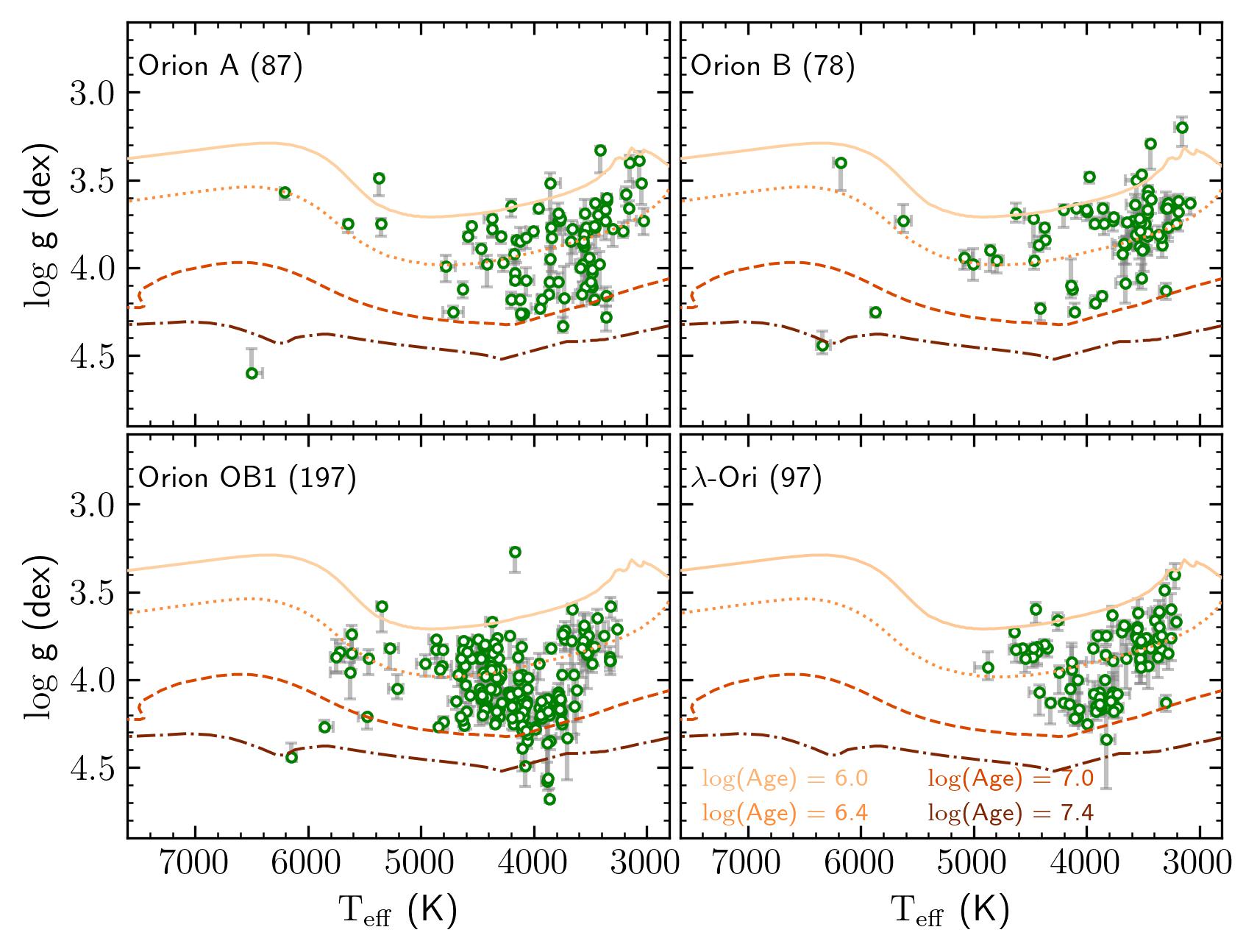}
    \caption{Kiel diagrams obtained with \tona\ for the 548 stars of Orion split into the four groups analyzed. As a reference, we include, as color lines, evolutionary models of \citet{parsec} for solar metallicity at $\log$(Age) = 
    6.0, 6.4, 6.8, and 7.2, corresponding roughly to 1, 2.5, 6.3, and 15 Myr, respectively. The numbers enclosed in the parentheses are the sources included in each stellar group.}
    \label{fig:kiel}
\end{figure*}

We discarded from our analysis five stars as their interquartile range in \teff\ is higher than 
500 K and six more with interquartile range in \vsini\ higher than 5 km s$^{-1}$, as we consider them inaccurate determinations. We report in Table \ref{tab:res} the atmospheric
parameters found for the remaining 548 young stars of Orion. In Figure \ref{fig:kiel}, we show the 
Kiel diagram (\logg\ vs \teff) for our sample. In general, the \teff\ and \logg\ values found 
with \tona\ lie in the expected range for young stellar objects as predicted by the evolutionary tracks of \cite{parsec}. 

For most of the objects in our sample, we determined \teff\ between 3200 and 
6000 K, corresponding approximately to spectral types M5/M4 to F9 \citep[e.g.,][]
{herczeg14}, although we found some stars with temperatures as high as 
$\sim$6400 K. \cite{adame24} found that \tona\ is capable of recovering reliable 
atmospheric parameters for stars with \teff\ below $\sim$6200 K. This behavior is 
linked to the theoretical grid of \cite{jonsson20}, as at higher \teff\ the number of 
spectral features becomes scarcer, complicating the fitting process and increasing the 
sensitivity to grid degeneracies.

Regarding \logg, our estimates show a bimodal distribution, with peaks around $\sim$3.8 and $\sim$4.2 dex, spanning a total range from approximately 3.3 to 4.6 dex. This bimodality may reflect slight differences in stellar ages among the four Orion subgroups analyzed, given that \logg\ serves as a proxy for stellar age. 

The \meta\ values obtained with \tona\ for the Orion stars are centered at 
$-$0.08$\pm$0.12 dex, which is consistent with the solar metal content within the 
errors. On the other hand, we found a mean \alfa\ for our sample around $-$0.03$\pm$0.06 dex, which is a value also consistent with the solar value, and it is well below the grid step of 0.25 dex. The metallicity and $\alpha$-elements enhancement estimations are in agreement with recent determinations of \cite{kos21}. However, these values should not be considered precise measurements; instead, they represent approximate global estimates from the synthetic models that best fit the observed spectra and are mainly suitable for first-order comparisons. To derive more accurate abundances, a detailed line-by-line analysis is necessary.

\begin{figure*}
    \centering
    \includegraphics[width=\textwidth]{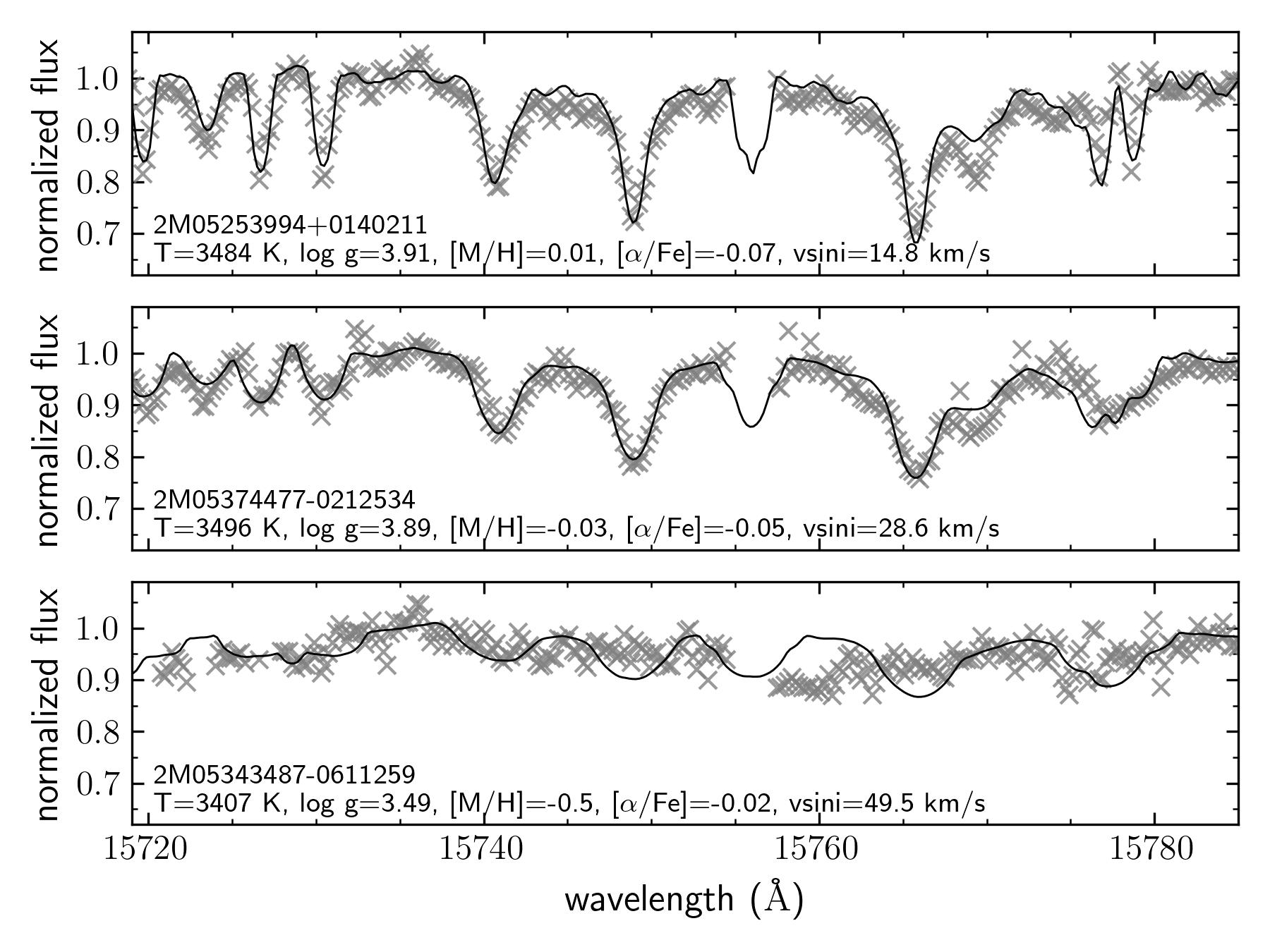}
    \caption{Small spectral window of three different observed spectra (gray crosses). In each panel, we overplot the synthetic spectra corresponding to the best-fit atmospheric parameters derived with \tona. The upper and middle panels show good-quality fits, while in the bottom panel the fit quality deteriorates as \tona\ approaches the upper \vsini\ limit of the grid.}
    \label{fig:qual}
\end{figure*}

Finally, we identified that 73\% of our sample (402 stars) have projected rotational 
velocities (\vsini) $\leq$ 35 km s$^{-1}$, with a mean value of 20 km s$^{-1}$. In 
contrast, a substantial subset of 98 stars exhibit \vsini\ values between 45 and 50 km 
s$^{-1}$. It is important to emphasize that atmospheric parameters for these high 
\vsini\ sources should be interpreted with extreme caution. First, edge effects may 
compromise the reliability of the results, as our analysis explores velocities only up 
to 50 km s$^{-1}$. {For instance, if a star intrinsically rotates faster than this limit, \tona\ will be unable to adequately broaden the synthetic spectrum to match the observed line depths. As a result, the code compensates by adjusting other parameters, such as \teff, \logg, or \meta, in an attempt to reproduce shallower lines, which can lead to anomalous or unphysical values.
This effect is not exclusive to \vsini; it applies to all parameters. Once the algorithm reaches the edge of the parameter space in any dimension, it tends to compensate by aggressively modifying the remaining parameters, thereby introducing potential biases. As we show in the next section, we found that the most objects with \vsini\ between 45 and 50 km s$^{-1}$ exhibit artificially low \meta\ values.} Second, increasing \vsini\ leads to significant line blending, which 
degrades our ability to accurately determine parameters, particularly those with 
subtler spectroscopic signatures, such as \logg\ and \meta. In these cases, our derived 
values should be considered as first-order approximations. {In Figure \ref{fig:qual}, we show a portion of the observed spectra overplotted with the corresponding best-fit synthetic spectra for three different Orion stars. These stars have similar values of \teff, \logg, \meta, and \alfa, but significantly different \vsini\ values, highlighting the challenges in characterizing fast rotators. The fits for the stars with moderate \vsini\ (upper and middle panels) show excellent agreement with the observations. In contrast, the star in the bottom panel, which has a higher rotational velocity, exhibits a noticeably poorer fit. This example illustrates how high rotation complicates spectral analysis and may lead to less reliable atmospheric parameters, particularly for \logg\ and \meta.}

\subsection{Comparing \tona\ parameters with literature}
We compared our stellar parameters with those from three different studies: The APOGEE Stellar Parameters and Chemical Abundances Pipeline in its version corresponding to DR17 \citep[ASPCAP DR17; ][]{dr17}, Apogee Net III \citep[ANet III;][]{anet3}, and Gaia DR3 \citep{gaia_dr3}. 

The ASPCAP Pipeline used a minimum $\chi^2$ statistic to compare observed APOGEE spectra to a set of model libraries in a two-step process that determined stellar parameters and abundances.

On the other hand, ANet III \citep{anet3}, and its predecessor \citep{olney20, anet2}, used a deep convolutional neural network to predict \teff, \logg, and \meta\  based on previously derived parameters (or “labels”) collected for APOGEE data and other stars in diverse studies. 

Finally, the Gaia DR3 \citep{gaia_dr3} team estimated \teff, \logg, \meta, radius, absolute 
magnitude $G$, distance and line-of-sight extinction from low-resolution optical $BP$/$RP$ 
spectra, apparent $G$ magnitude and parallax. They employed a Markov Chain Monte Carlo (MCMC) 
algorithm in three different steps to derive the parameters mentioned above in a self-
consistent manner from four grids of synthetic spectra and a set of PARSEC COLIBRI isochrones 
\citep{Tang14, chen15, pastorelli20}.

Among the three pipelines mentioned above, ANet III focuses more specifically on the characterization of young stars, producing parameters that are more adequate for a direct comparison with our work.

\begin{figure*}
    \centering
    \includegraphics[width=\textwidth]{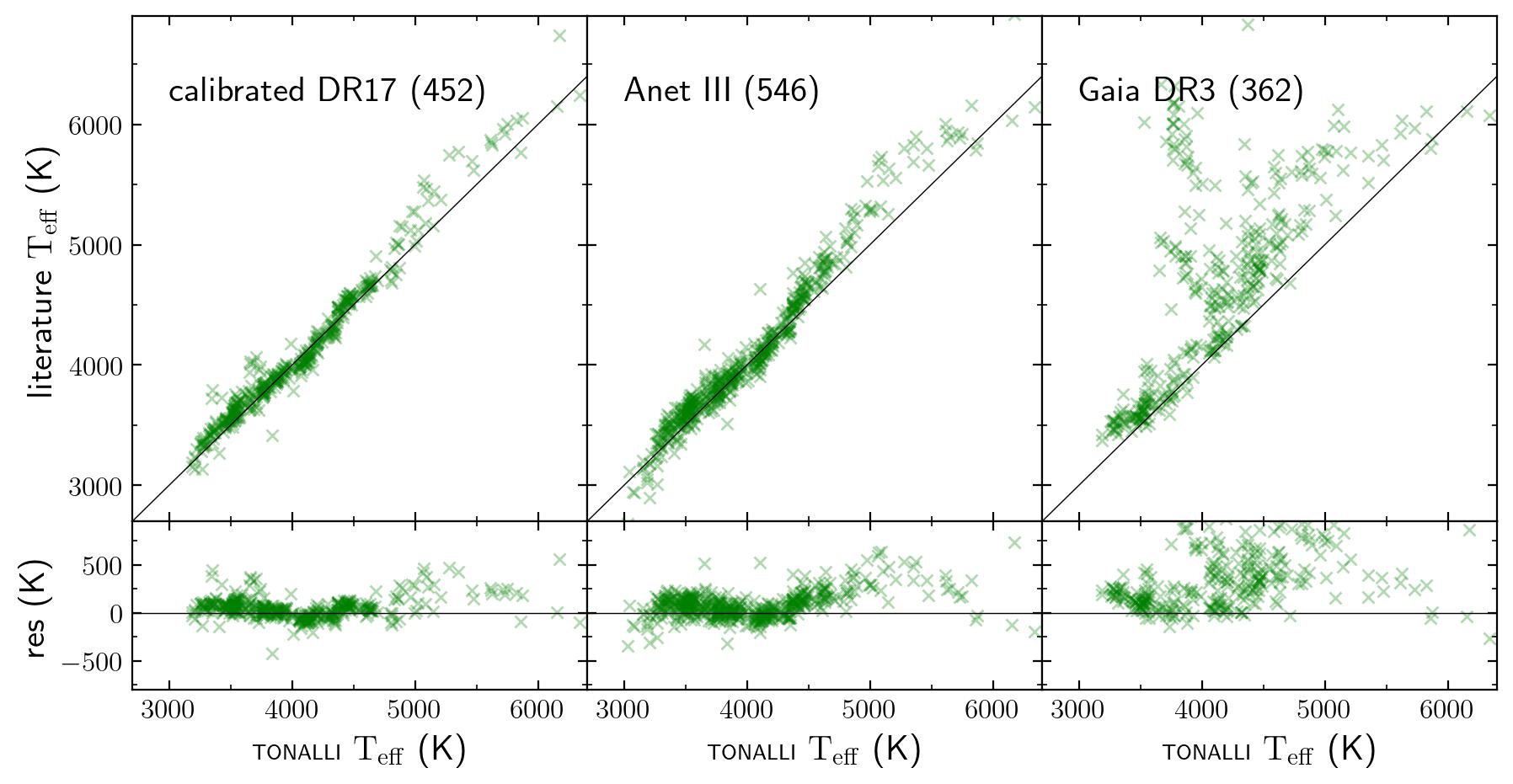}
    \caption{Comparison of the \tona\ \teff\ with the calibrated estimates of ASPCAP DR17 \citep{dr17}, ANet III \citep{anet3} and Gaia DR3 \citep{gaia_dr3}. The number within the parenthesis is the stars included in each comparison. The lower panels are the differences between the literature and our estimates.}
    \label{fig:temps}
\end{figure*}

In Figure \ref{fig:temps}, we compare our temperature estimates with the studies mentioned above. We have good agreement with ASPCAP DR17 and ANet III as we found a mean difference (literature - ours) of 62$\pm$213 K and 92$\pm$238 K, respectively. {Both comparisons reveal a similar trend: hotter stars exhibit larger discrepancies between studies. This divergence becomes noticeable for sources with \tona\ \teff\ above 5000 K, where the mean temperature difference reaches 190$\pm$170 K and 290$\pm$213 K for ASPCAP DR17 and ANet III, respectively. In contrast, for stars cooler than 5000 K, the differences are significantly smaller, just 37$\pm$90 K and 66$\pm$112 K. This behavior may stem from the fact that hotter stars have fewer spectral features in the $H$-band, making temperature determinations more challenging. For \tona, temperature estimates become less reliable beyond $\sim$6000 K \citep[]{lopezval24}.
The comparison with ASPCAP DR17 shows smaller discrepancies, likely due to the use of the same model atmospheres and observational data. Nevertheless, differences in continuum normalization and parameter determination methods may still introduce variations (though these remain within typical uncertainties). In contrast, ANet III relies on a machine learning approach based on the spectroscopic labels of its training set, which may introduce a systematic offset that contributes to the observed differences.} 

However, some Gaia DR3 temperatures are overestimated, in extreme cases up to 2000 K. This effect might be due to the degeneracy of the evolutionary tracks for giants and young stars used in the Gaia determination pipeline.   

\begin{figure*}
    \centering
    \includegraphics[width=\textwidth]{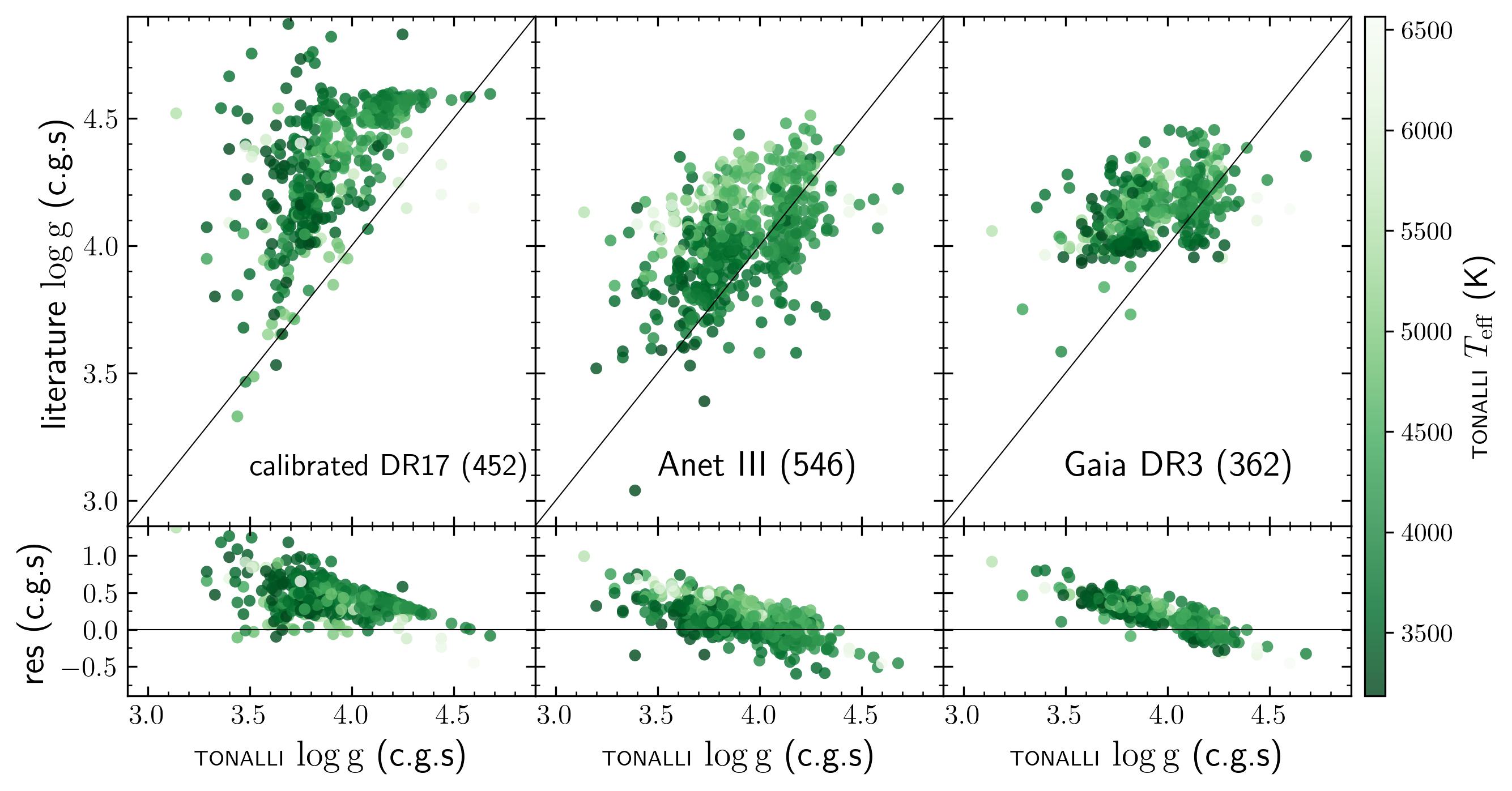}
    \caption{The same as Fig. \ref{fig:temps} but for \logg. The points are color-coded by \tona\ \teff. }
    \label{fig:gras}
\end{figure*}

Regarding \logg, we observe systematic trends in the residuals shown in Figure~\ref{fig:gras}. Across all three panels, the trend is consistent: at lower \tona\ \logg\ values, literature estimates tend to be higher by up to 1.0 dex. This discrepancy decreases as \logg\ approaches $\sim$4.1 dex, at which point the trend inverts, with \tona\ values becoming higher, up to 0.5 dex, than those reported in the literature. Surface gravity is arguably the most challenging atmospheric parameter to determine, as its influence on spectral lines overlaps with that of other parameters but with subtler effects. Notably, several ASPCAP DR17 \logg\ values cluster at 4.5 dex, which appears slightly too high for young, low-mass stars. Notably, several ASPCAP DR17 \logg\ values cluster near 4.5 dex, which would be too high for pre-main sequence stars; however, it is worth noticing that ASPCAP DR17 is optimized for main- and post-main sequence targets. In contrast, Gaia DR3 estimates are more uniformly distributed around $\sim$4.1 dex, yet they remain on average 0.20 dex higher than our determinations and, more importantly, higher than expected for the Orion age. Despite the observed trends, our best agreement is with ANet III, where our \logg\ values are on average 0.13 dex higher. {The points in Figure \ref{fig:gras} are color-coded by \tona\ \teff; however, no strong trend is observed. There is a slight indication that stars with higher \logg\ in ANet III and Gaia (middle and right panels) tend to be hotter in our analysis.}

\begin{figure*}
    \centering
    \includegraphics[width=\textwidth]{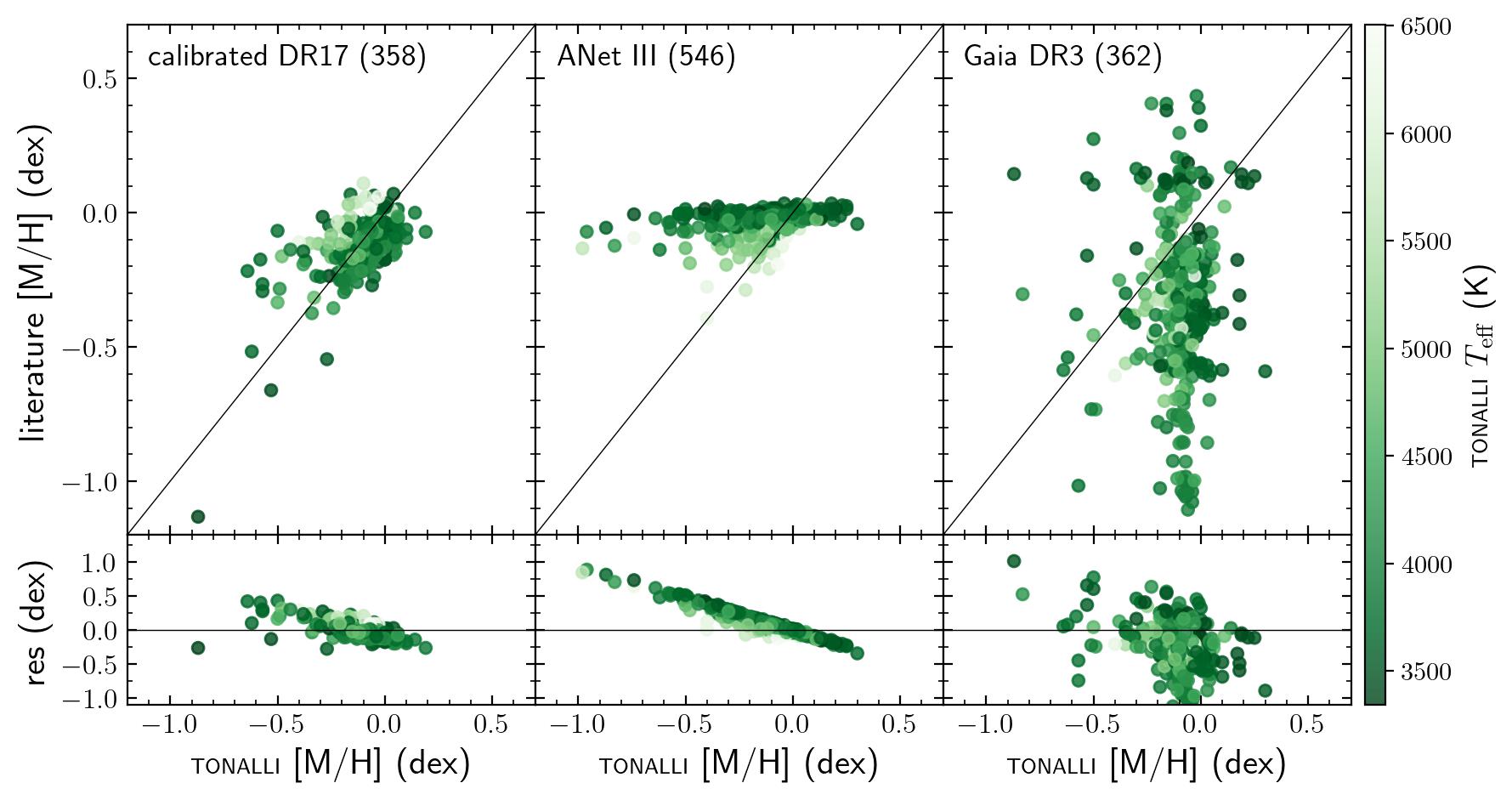}
    \caption{The same as Fig. \ref{fig:temps} but for \meta. The points are color-coded by \tona\ \teff.}
    \label{fig:meta}
\end{figure*}

Finally, the \meta\ values exhibit a trend similar to that of \logg, as shown in Figure~\ref{fig:meta}. We find the best agreement between our \meta\ values and the ASPCAP DR17 estimates, with our measurements being, on average, 0.02$\pm$0.10 dex higher. In contrast, the \meta\ values from ANet III for Orion stars are more tightly constrained around solar metallicity (0.0$\pm$0.2 dex), likely due to the spectral ``labels'' used to train their neural network. Meanwhile, the Gaia DR3 metallicity estimates show substantial scatter across the plot. This high dispersion is likely due to the low spectral resolution of the $BP/RP$ spectra from which the \meta\ values were derived. Additionally, the Gaia metallicity estimates appear systematically too low for the types of sources studied here. {Similarly to Figure \ref{fig:gras}, we color-coded the comparisons in Figure \ref{fig:meta} using \tona\ \teff. We found a mild trend in the ANet III panel, where lower \meta\ values appear to be associated with hotter stars in our analysis. This behavior was previously reported for earlier versions of the ANet algorithm, which tended to predict lower metallicities for hotter stars \citep{olney20}. Although newer versions of the algorithm have improved in this regard, our results suggest that this issue may still persist to some extent in young stars. }

The atmospheric parameters derived with \tona\ for the Orion stars span \teff\ and \logg\ ranges consistent with expectations from evolutionary models. However, 22 sources exhibit metallicities below $-0.50$ dex, which may be incompatible with the young age of Orion and the chemical properties of the thin disk stellar population. Among these sources, nineteen ($\sim$86\%) have projected rotational velocities (\vsini) exceeding 47 km s$^{-1}$. In these cases, the actual stellar rotation may exceed the maximum \vsini\ value allowed in \tona\ (50 km s$^{-1}$). {As we mentioned before, } when this upper limit is reached, the algorithm struggles to reproduce the observed spectral line broadening. As a result, it compensates by artificially lowering the \meta\ in an attempt to match the shallow spectral features, leading to anomalously low values. As previously noted, the atmospheric parameters of sources 
with \vsini\ $\geq$ 45 km s$^{-1}$ should be considered with caution and treated as first-order approaches. For this reason, the low \meta\ values in these high \vsini\ stars are likely outliers rather than reliable measurements and should be discarded from further analysis. 

\input{space}

\section{Chemical abundance determination}
To determine the chemical abundances of our sample, we used the atmospheric parameters obtained from \tona, the Brussels automated code to characterize high-accuracy spectra \citep[{\tt BACCHUS};][]{masseron16}, and the reduced spectra of APOGEE DR17.

\bacc, computes on-the-fly synthetic spectra for a range of abundances that compares to 
observations on a line-by-line basis. The version of \bacc\ we used relies on the v15.1 of the 
1D LTE radiative transfer code {\tt TURBOSPECTRUM} \citep{turbo}. To generate the synthetic spectra 
we employed a MARCS model atmosphere grid \citep{jonsson20}, the atomic line list of 
\cite{smith21}, and the molecular line lists for CO, CN, and OH of \cite{li15}, 
\cite{sneden14} and \cite{brooke16}, respectively. 

Additionally, to compute a synthetic spectrum model, {\tt TURBOSPECTRUM} needs to be provided with (or compute) the basic atmospheric 
parameters (\teff, \logg, \meta, \alfa), microturbulence velocity, a convolution parameter, 
which includes the instrumental and rotational broadening, and a standard solar composition, which in our case was that of \cite*{grevese07}. 
To derive chemical abundances, \bacc\ compared, within a 30\AA\ spectral window, the synthetic 
and observed spectra through four different methods, namely, \emph{chi2}, \emph{syn}, 
\emph{eqw}, and \emph{int} to find the best A(X)\footnote{A(X) = $\log$ ($n_{\rm X}$/$n_{\rm 
H}$) + 12} abundance for each method. In this study, we used the \emph{chi2} method, which minimizes the squared differences between observed and synthetic spectra. This method is the most balanced among the four. 

For each star in our sample, we used the epoch-combined 1D APOGEE spectrum\footnote{We used the 
apStar spectra downloaded from the SDSS-IV science archive web app (SAW)},  the \tona\ \teff, 
\logg, \meta, and initially we used the line lists reported in \cite{lopezval24} and \cite{vinicius24} that allowed us to access to twelve light elements: C, Na, Mg, Al, Si, K, Ca, Ti, V, Cr, Mn, and Fe. 

We left the microturbulence velocity and the convolution parameter of \bacc\ to vary freely. 
However, we initialized the convolution parameter to be equal to \vsini\ value found in \tona, something of great importance, as with this, the convergence effectiveness of \bacc\ improved. 
The outcome of this process was the chemical abundance, together with a quality flag, SNR, and 
equivalent width for each atomic line in the line list {as well as a value of microturbulence and convolution parameter for each star. } 

\subsection{Final abundance and error}
To compute the final abundances and their uncertainties, we used the best-fit parameters derived with \tona. However, since these parameters have associated uncertainties, we adopted a Monte Carlo approach to propagate their effects on the abundance determinations.
Specifically, for each star, we repeated the abundance determination procedure 100 times, each time perturbing the atmospheric parameters (\teff, \logg, \meta, and \vsini) by randomly sampling values from a uniform distribution within their respective interquartile ranges as determined by \tona. In each iteration, we ran \bacc\ to compute the abundances for all available atomic lines, but selecting only lines with a good-fit flag and signal-to-noise ratio greater than 100 to ensure high-quality measurements.
As a result, for each star and each selected atomic line, we obtained a distribution of 100 abundance values. From this distribution, we adopted the median as the final abundance estimate and defined the uncertainty as the range between the minimum and maximum values found across the 100 iterations.

This procedure to obtain the uncertainties on the chemical abundances was time-consuming as \bacc\ is not parallelized. To solve this, we used the newly created GRID UNAM system, which allows us to run {\tt BACCHUS} in a pseudo-parallel way to reduce the computing time by occupying as many threads available at a given time in a server grid distributed among various institutions. The Grid UNAM is a high-performance computing infrastructure that brings together various units within the National Autonomous University of Mexico (UNAM) to enable the efficient and collaborative use of computational and human resources. It operates through a 10 Gbps network that connects distributed computing and storage systems, offering both high-performance and high-throughput computing services. Its heterogeneous infrastructure includes powerful computing clusters and large-scale data storage supported by the MinIO\footnote{\url{https://min.io/}} platform, widely used for handling workloads in machine learning, data analytics, and scientific applications.

Recently, some studies have found trends between determined chemical abundances and the stellar parameters, mainly temperature \citep[e.g., ][]{kos21,vinicius24,kos25}, derived from spectra of the same or similar spectral resolution. In \cite{kos25}, there is an extended discussion on the trends they found between \teff\ and abundance value in stellar clusters. They argue that trends between abundance and another stellar parameter like gravity or \vsini\ might be the result of the relation of these parameters with \teff, and that those trends disappear once detrending is applied to the abundances. \cite{kos25} observed two main behaviours in their analysis, thanks to the fact that they included a wide range of stars (e.g., giants and dwarfs) of different ages. They found a monotonic trend between 4800 and $\sim$6800 K, similar to the behaviour found by  \cite{vinicius24} in FGK dwarfs stars of the Pleiades. The second trend found by \cite{kos25} was for lower temperature stars, and it is steeper than the first, even showing a sharp upturn at $\sim$3800 K. Although the origin of the \teff-abundance trends is not clear, there might be multifactorial causes related to physical processes (e.g., atomic diffusion) and processing issues (e.g., continuum fitting, linelists, atmospheric models) as suggested by \cite{kos25}. 
Unveiling the nature of these trends is beyond the scope of this work; however, we favoured the interpretation that these trends are primarily driven by limitations in the atmospheric models and linelists. {We refer the reader to Section 4 of \citet{kos25} for a more detailed discussion on the origin of these trends.} As a practical solution, we identified 
and excluded from our analysis those atomic lines that exhibit a significant systematic 
trend as a function of either \teff\ or \logg.

\begin{figure}
    \centering
    \includegraphics[width=\columnwidth]{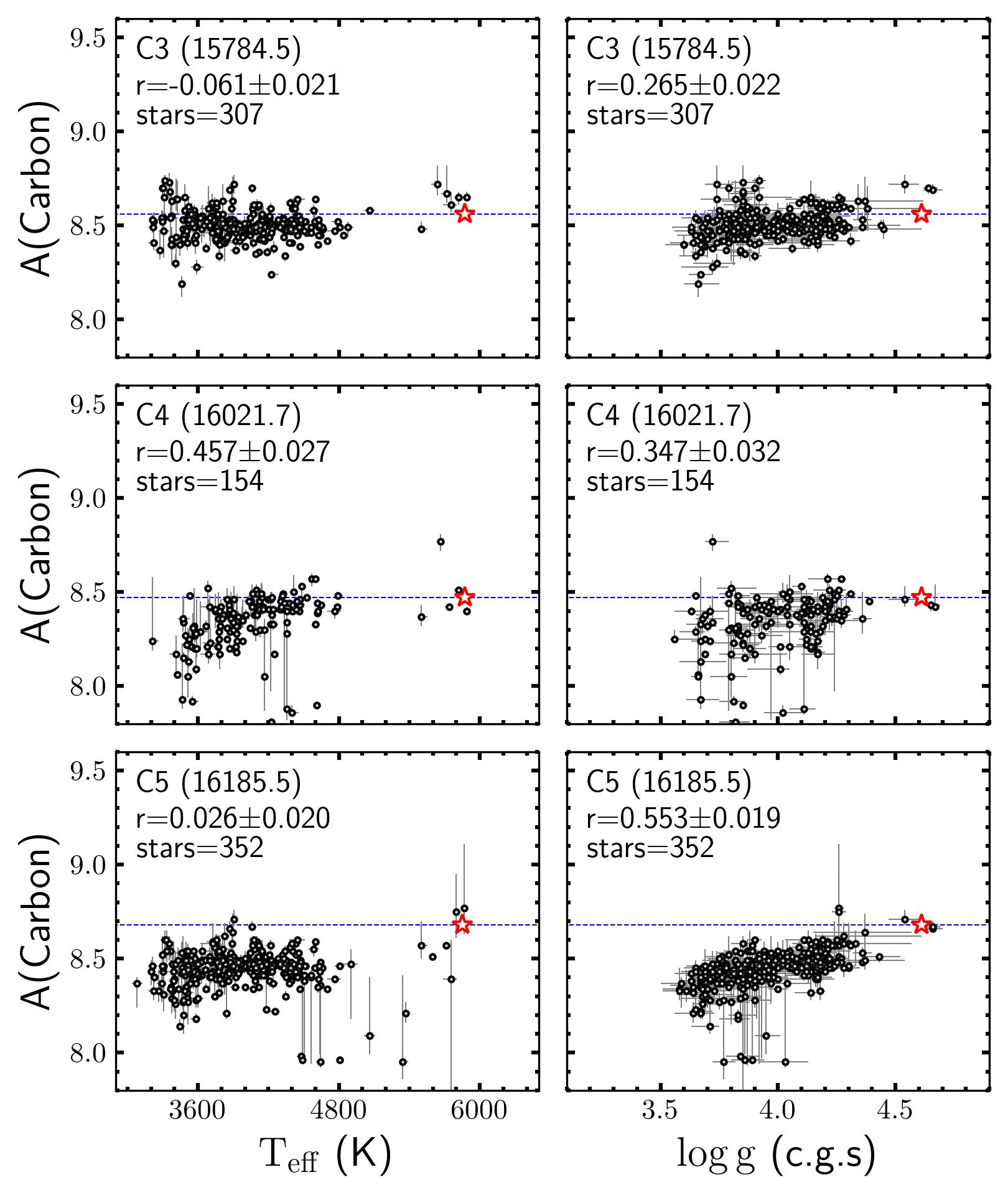}
    \caption{Logarithmic abundance values found with \bacc\ for three different carbon lines (C3, C4, C5) are shown as a function of \teff\ (left) and \logg\ (right). The number in parentheses indicates the wavelength of each line, while $r$ represents the mean Spearman correlation coefficient calculated for the corresponding dataset. The red star and dashed line indicate the solar abundance associated with the line under analysis. Based on our selection criteria (see text), only the line identified as C3 was ultimately included in our analysis.}
    \label{fig:lines}
\end{figure}

We computed the Spearman correlation coefficient ($r$) to discard those lines that 
exhibited a moderate to high linear correlation between abundance and \teff\ and 
abundance and \logg. To reinforce the significance of the $r$ coefficient computed, we 
randomly varied the abundance value within its error and recomputed Pearson 
coefficients. After 300 iterations, we computed a mean $r$ value and the standard 
deviation to assess the correlation of the examined atomic line with \teff\ and \logg. 
We then selected lines with a weak linear correlation, corresponding to $r\pm\sigma_r$ 
values within the $\pm$0.35 range. 
{In Figure \ref{fig:lines}, as an example, we present the abundance values derived with \bacc\ for three different carbon lines, plotted as a function of \teff\ (left column) and \logg\ (right column). Among the three lines, only the one identified as C3 (top panels), located at approximately 15784 \AA, was included in our final analysis, as it exhibits $r$ below 0.35 with both \teff\ and \logg. In contrast, the C4 and C5 lines failed to meet this criterion, with only one of their coefficients falling below the threshold, and were therefore excluded from further analysis.} 
We found 19 atomic lines of 6 elements (C, Mg, Si, K, Ti, and Fe) that met our $r$ 
value requirement, and those that we included in our subsequent analysis. In 
Table~\ref{tab:lines}, we present the basic information of the atomic lines used in this work. It is important to mention that the selection threshold of the $r$ coefficients is set arbitrarily, and if we relaxed it, more lines would meet the criteria; however, abundance trends might appear in our results. 

{The exclusion of tendentious lines enhances the robustness of our abundance determinations, reducing the likelihood that observed variations arise from the line list itself. This approach enables more reliable comparisons across different groups or stellar members, minimizing spurious trends with effective temperature or surface gravity. While all abundance analyses inherently depend on the adopted model atmospheres and line lists, our methodology is designed to mitigate such dependencies. Additionally, although some elements can exhibit departures from LTE that introduce trends with \teff\ and \logg, these effects are expected to be smaller in the near-infrared, particularly for cool dwarf stars with near-solar metallicity, than at optical wavelengths \citep[e.g.,][]{osorio20, olander21}. To quantify this, we used the publicly available MPIA non-LTE corrections service (\url{https://gemini-web.mpia.de/}), which compiles results from infrared studies of Mg, Si, Ti, and Fe \citep[e.g.,][]{bergemann12, bergemann13, bergemann15}. For these elements, the corrections are negligible (typically <0.02 dex). For potassium, the corrections are below 0.1 dex \citep{osorio20}, which is comparable to the mean uncertainty in our K abundance (0.06 dex)}

\subsubsection{[X/H] abundance ratios}
We computed [X/H]\footnote{The [X/H] = A(X)$_{\rm star}$ - A(X)$_\odot$, where the A(X) are the \bacc\ abundances found} ratio for each selected line using the solar abundance obtained, in the same fashion as our sample, from the APOGEE DR17 Vesta spectrum (Table \ref{tab:lines}). We used the 1D solar atmospheric parameters reported by \cite{adame24}.  If we had more than one selected line from the same element, we computed the mean value and propagated the individual errors to obtain a final uncertainty. 
 
We did not measure the abundances for all the stars of our sample for three main reasons: i) the temperature of the star (too hot or cold) interfered to measure reliably some of the selected atomic lines; ii) there were cases where some spectral regions presenting data reduction issues, and iii) the \vsini\ of the star was high ($>$ 30 km s$^{-1}$), implying a higher degree of line blending. We also did not report [X/H] abundance estimations higher than 0.5 dex and lower than -0.75 dex, as we considered them as values incompatible with the stellar population we are working with. We measured the abundance of two or more elements for 340 stars ($\sim$62\%) of our sample, that we report along with their respective errors in Table~\ref{tab:res}.
{Additionally, we examined the relationship between the microturbulence values derived with \bacc\ and the mean elemental abundances. No strong correlations were found, with Pearson coefficients ranging from $r = -0.29$ to $-0.07$ across all elements analyzed. We also investigated the potential correlation between \vsini\ and the \bacc\ microturbulence values, again finding no significant trend. The median \bacc\ microturbulence in our sample is 1.1 km s$^{-1}$, and individual values for each star are reported in Table~\ref{tab:res}}
\input{resul}

\input{tab_lines}


\input{tab_groups}

\begin{figure*}
    \centering
    \includegraphics[width=\textwidth]{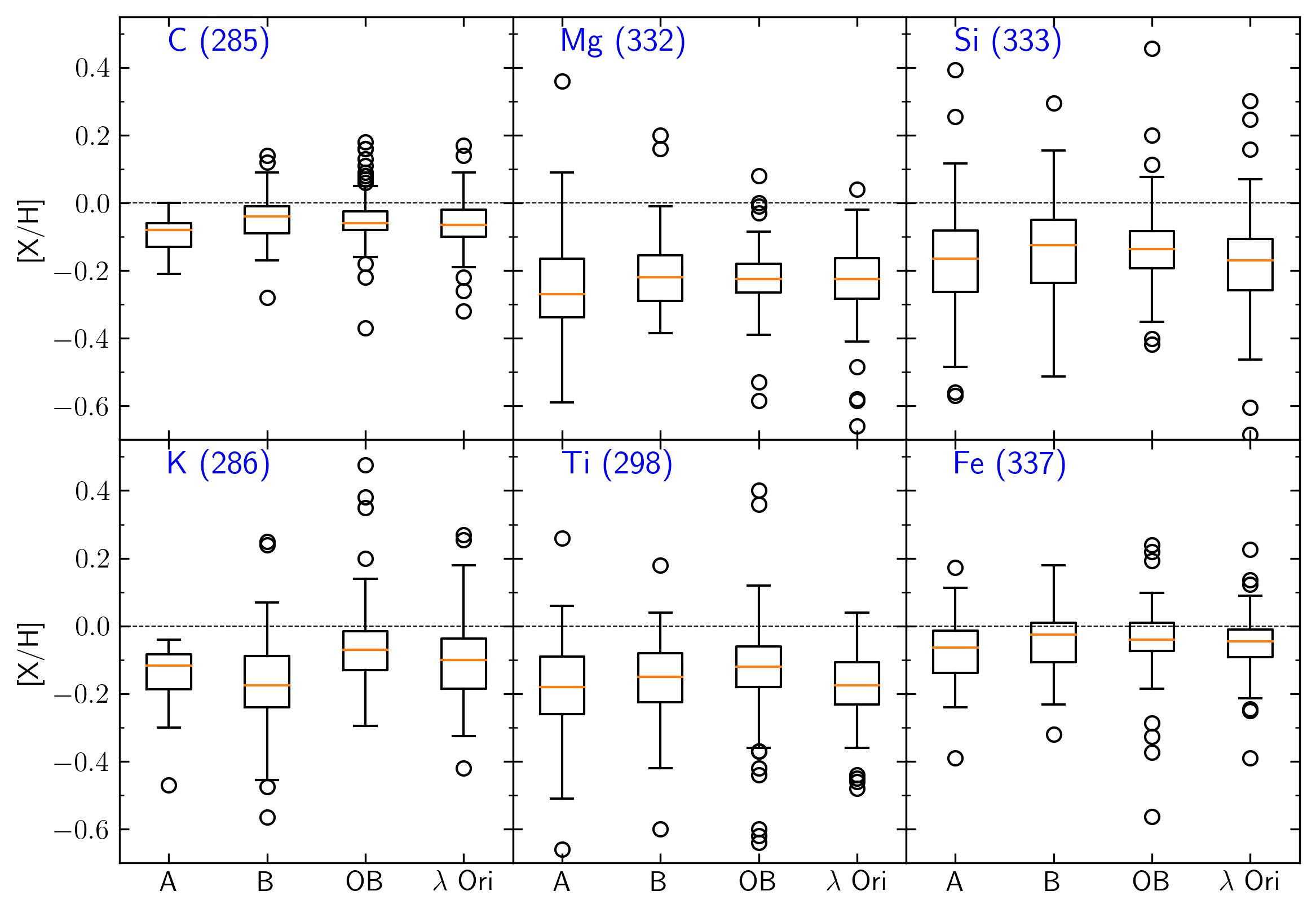}
    \caption{Boxplot of the [X/H] abundance ratios derived in this work for each Orion group. The box encloses data between the inter-quartile range (IQR) while the whiskers extend from the box to 1.5$\times$IQR. The orange horizontal line is the median value. The total number of stars for each element is within parentheses. Our determinations suggest that Orion is chemically homogeneous,  as supported by the low MAD shown in Table \ref{tab:groups}.}
    \label{fig:vio}
\end{figure*}

\section{Chemical abundances of Orion}
According to our [X/H] abundance ratios, the whole Orion complex presented sub-solar abundances for all six elements studied. {This result aligns with the presence of a negative vertical metallicity gradient observed in the solar neighborhood using samples of giant and hot stars \citep[e.g.,][]{hawkins23,hackshaw24}. The Orion star-forming region, located approximately 100–180 pc below the Galactic midplane \citep[e.g.,][]{zucker20,gross21}, lies along the Radcliffe Wave \citep{alves20}, where similar trends have been reported. Furthermore, our findings agree with Galactic disk metallicity maps, which show average metallicities of about –0.10 to –0.05 dex at the position of our sample \citep[e.g.,][]{poggio22,martinez-medina25}. While most of the aforementioned studies focus on evolved or massive stars, the agreement with our results based on young, cool stars is promising. It suggests that this type of analysis could be extended to other star-forming regions to further investigate the radial and vertical metallicity gradients in the Galactic disk.}

Figure \ref{fig:vio} shows box plots of our [X/H] values to visually analyze the 
abundance ratios in the different Orion groups. As previously mentioned, the median 
abundance ratios for all elements are sub-solar. Despite this, most of them show a 
dispersion that still makes these ratios consistent with solar abundances. 

In terms of comparing the different groups, although there might be some differences 
between median abundance values, this was not significant, and all of them were 
consistent within their corresponding MAD, suggesting that the Orion complex was 
chemically homogeneous in all six elements analysed here (see Table~\ref{tab:groups}). 
Our results agreed with \cite{kos21}, where they also found slightly sub-solar 
abundances for Orion members and chemical homogeneity.

Comparing chemical abundances across different studies presents inherent challenges, as methodologies, atmospheric models, and standard solar scales often differ. In many cases, a calibration to a common framework is necessary to ensure meaningful comparisons \citep[e.g.,][]{hinkel14}. Nonetheless, relative comparisons can still provide valuable insight.

To place the Orion members in context, we used the sample from \citet{lopezval24} as a reference for the chemical abundances of the solar neighborhood. This comparison sample includes G-, K-, and M-type stars with effective temperatures between 3300 and 6000 K, all located within 100 pc of the Sun. We adopted the atmospheric parameters reported in that study and applied the same abundance determination methodology used in this work to ensure consistency between the two datasets.

The nearby sample is highly diverse, comprising stars with a wide range of masses and ages. To ensure a homogeneous comparison between Orion and the solar neighborhood, we divided both samples into two \teff\ bins: stars with \teff\ $\leq$ 4000 K (M-type stars), and stars with temperatures between 4000 and 5000 K (K-type stars). We also restricted the comparison to stars with [Fe/H] values within $\pm$0.5 dex. These two temperature bins encompass 93\% of the Orion stars for which \bacc\ chemical abundances were determined.

We compared the Orion [X/H] abundances to those of the solar neighborhood and found 
that the median values are quite similar between both samples in both \teff\ bins (see 
Table~\ref{tab:ism_comp}). This agrees with previous studies that report remarkable 
chemical homogeneity across the solar neighborhood, based on both nebular and stellar 
    tracers \citep{Esteban:22,mendez22,Ritchey:23}.

However, Kolmogorov–Smirnov (KS) tests indicate that the two populations are 
statistically different, with very low $\mathcal{P}$(KS) values. Since the nearby 
sample contains many more stars than the Orion sample, we tested whether this 
difference in sample size might be responsible for the low KS probabilities. We 
performed a bootstrap procedure, randomly drawing sub-samples from the nearby 
population with the same number of stars as the Orion sample, and computed the KS test 
10,000 times. We report the median of the resulting $\mathcal{P}$(KS) distribution in 
Table~\ref{tab:ism_comp}. The low median $\mathcal{P}$(KS) values persisted, 
reinforcing our initial result.

Since both samples were restricted in \teff\ and [Fe/H], the observed differences might 
be, at least partially, attributed to differences in \logg, which can be interpreted as 
a proxy for age. For M-type stars, we found median \logg\ values (and MAD) of 
3.86$\pm$0.17 for Orion and 4.78$\pm$0.05 for the main-sequence stars. This difference 
is smaller for K-type stars, where the medians were 4.06$\pm$0.13 and 4.68$\pm$0.04, 
respectively.

Additionally, using our Mg, Si, and Ti abundance determinations, we computed 
[$\alpha$/Fe] for both samples. {We define [$\alpha$/Fe] as a weighted average of the [Mg/Fe], [Si/Fe], and [Ti/Fe] abundance ratios. In this calculation, more weight is given to the measurements with smaller uncertainties, so that the most precise values contribute more significantly to the final [$\alpha$/Fe] estimate. The resulting [$\alpha$/Fe] values and their corresponding uncertainties, calculated as the error in the weighted mean, are listed in Table \ref{tab:res}.}

For the M-type stars, the median [$\alpha$/Fe] 
{were $-0.14\pm0.05$ and $-0.04\pm0.06$} for Orion and the main-sequence 
samples, respectively. Similar values were found for the K-type stars: {\bf $-0.12\pm0.03$} 
for Orion and {\bf $-0.05\pm0.03$} for the main-sequence stars.

\begin{figure}
    \centering
    \includegraphics[width=\columnwidth]{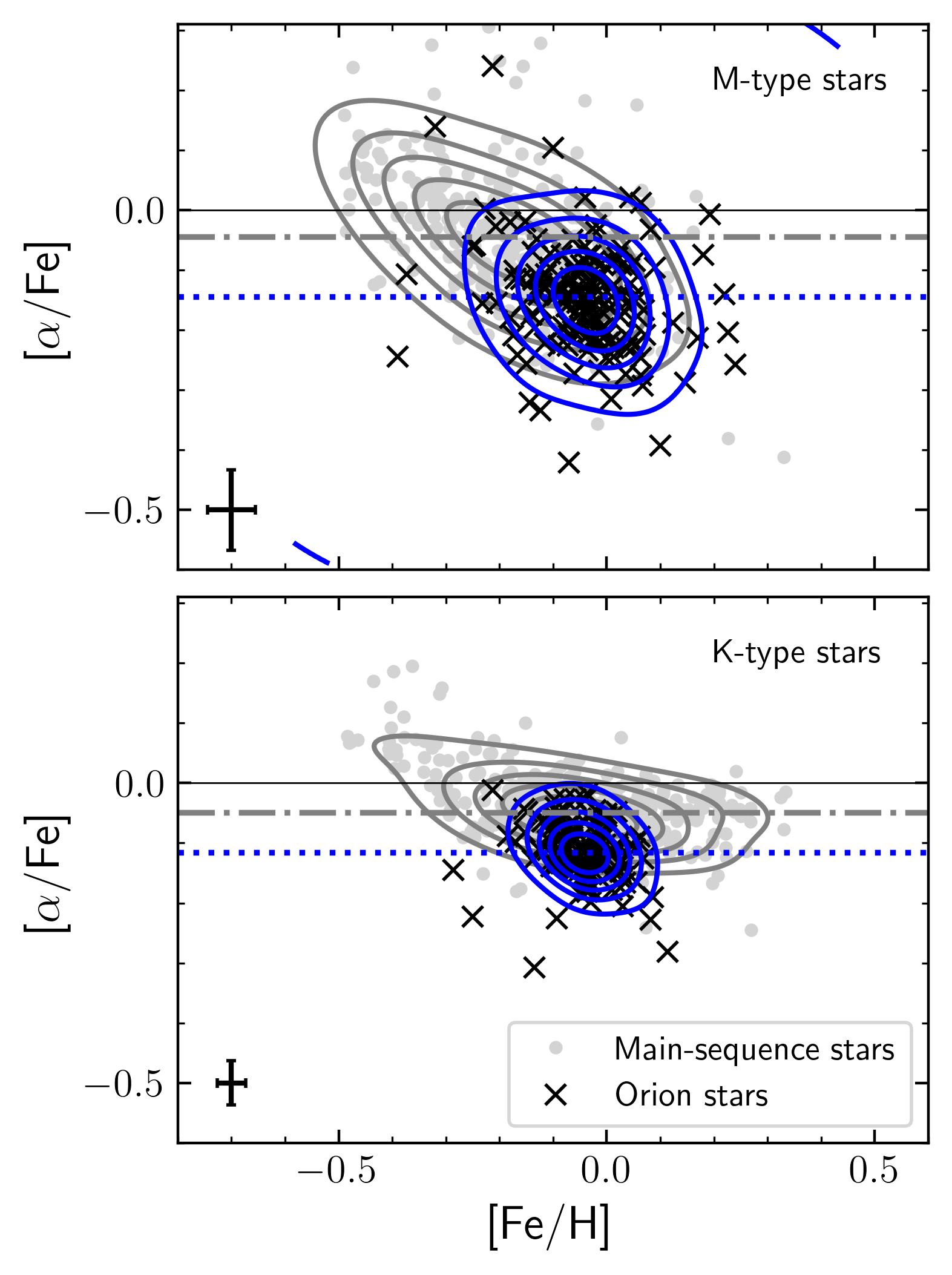}
    \caption{Tinsley-Wallerstein diagram for the Orion members (black crosses) and the main-sequence stars (gray points) of \citet{lopezval24} using the abundances determined in this study for M-type stars (upper panel) and K-type stars (bottom panel). The [$\alpha$/Fe] {ratio corresponds to the weighted average} of our [Mg/Fe], [Si/Fe], and [Ti/Fe]. The dotted and dashdotted lines represent the median value of [$\alpha$/Fe] for the Orion and main-sequence stars, respectively. We also included number density contours for Orion (blue) and the main-sequence stars (gray). {Error bars in the lower-left corner of each panel represent the median uncertainties in the [Fe/H] and [$\alpha$/Fe] for Orion stars.}}
    \label{fig:alfa}
\end{figure}

Figure~\ref{fig:alfa} shows the [Fe/H]–[$\alpha$/Fe] (Tinsley–Wallerstein) diagram for 
both populations. The abundances of the main-sequence stars fall within the expected 
locus of thin disk stars \citep[e.g.,][]{neves09}. The Orion members, on the other 
hand, occupy a more constrained region in this diagram and appear to be shifted toward 
lower [$\alpha$/Fe] and sub-solar [Fe/H] values. This may suggest that the interstellar 
medium from which these younger stars formed has already been enriched by the stellar 
evolution of older stellar generations.
From the evolutionary perspective, the chemical abundance of the ISM is a better 
comparison to the abundance of young stars, as the difference in the evolutionary stage 
is smaller than with the main-sequence stars. Moreover, the main-sequence stars of 
\cite{lopezval24} are located at different distances and positions in the Galaxy, something that might also introduce uncertainties. 

\cite{mendez22} analyzed the radial abundance gradients of He, C, N, O, Ne, S, Cl, and 
Ar in the Galaxy using optical spectra of 42 H {\sc ii} regions, including M42 (the Orion 
Nebula). They determined a carbon 
abundance of M42 of 8.34$\pm$0.02, which translates to -0.05$\pm$0.05, -0.12$\pm$0.04, 
and  -0.17$\pm$0.09 if the standard solar composition of \cite*{grevese07}, \cite{asplund21}, or \cite{l25} is adopted. The median [C/H] we have determined for the whole Orion complex is -0.06$\pm$0.03, which agrees, when we take into account the errors, with the three different values of M42. 

It is important to note that the C/H determination by \citet{mendez22} in the Orion Nebula is based on the C\,\textsc{ii} $\lambda4267$ recombination line (RL), whose ratio with H$\beta$ is virtually insensitive to the physical conditions of the gas. In contrast, the collisionally excited lines (CELs) C\,\textsc{iii}] $\lambda\lambda1907,1909$, measured in the UV by \citet{Walter:92}, are highly sensitive to the electron temperature ($T_e$) of the gas, and yield a C/H abundance approximately $\sim$0.4 dex lower \citep{Esteban:04}. The CEL-nebular calculations are in tension with the stellar abundances reported in this work.

The longstanding discrepancy between abundances derived from CELs and RLs has been widely debated in the literature \citep{MendezDelgado:2023a,MendezDelgado:2024a,Chen:2023}. Determining which of these two methods yields more accurate chemical abundances is critical to our understanding of the formation and chemical evolution of galaxies \citep{Curti:2025}. The stellar abundance measurements carried out by our group may offer crucial insight into this issue, providing an independent tracer of the true chemical content. The good agreement between RL-based carbon abundances and those derived from stellar objects supports the view that RLs may provide more reliable abundance estimates \citep{MendezDelgado:2023a}, possibly indicating that internal temperature fluctuations within the ionized gas \citep{Peimbert:67}, or other physical processes, may systematically bias CEL-based nebular abundances downward.

The Orion nebula, or M42, is spatially located around the Orion A group, thus, the stars of this group are a better test bed to sense the composition differences between the young stars and the H {\sc ii} region. In Figure \ref{fig:vio} (and Table \ref{tab:groups}) there is a hint that members of the Orion A group possess a slightly lower carbon abundance than stars in the other three groups, aligning with the sub-solar abundances reported by \cite{mendez22} for M42. The median [C/H] abundance for Orion A is -0.08$\pm$0.03, the lowest in absolute value, although not significant if errors are considered, among all the Orion groups. This comparison represents a valuable starting point for future studies combining the APOGEE-2 data with the SDSS-V Local Volume Mapper, which will provide more and better galactic ISM spectra. As we showed, the combination of both surveys gives us the possibility to do direct comparisons of the composition of gas and young stars for a specific region, such as in the case of Orion, and helps us to constrain Galactic chemical evolution models. Additionally, we can investigate gradients within star-forming regions as well as within the Galaxy through the characterization of young stars (I. Hern\'andez-Aburto in prep.), and connect this with the previous determinations made through the composition of the interstellar medium.  

The analysis of multiple chemical elements in stellar parameters is a rapidly growing area of research, not only within the stellar community but also in the context of nebular studies. As demonstrated by \citet{MendezDelgado:2024b}, comparing stellar Fe-based metallicities with nebular O-based metallicities, as has been done by some authors using low-resolution spectra \citep{Bresolin:2016}, is not only insufficient, but may also introduce systematic biases in the interpretation of chemical abundances, particularly in the broader context of galaxy evolution.

\input{ism_ab}

\section{Summary and remark conclusions}

We determined atmospheric parameters (\teff, \logg, \meta, \alfa, and \vsini) for a sample of 548 young stars in the Orion complex, using the spectral analysis code \tona\ \citep{adame24}. These stars, located in four subgroups (A, B, OB, and $\lambda$ Ori), were selected based on the absence of infrared excess, identified via their position in the $J$--($K-W3$) color–magnitude diagram. This selection minimizes the impact of continuum veiling on our parameter estimates. The atmospheric parameters derived with \tona, particularly \teff\ and \logg, are consistent with evolutionary models for stars at the age of Orion.

Using these parameters, we then applied the \bacc\ code \citep{masseron16} to determine the chemical abundances of C, Mg, Si, K, Ti, and Fe. We obtained reliable abundances for at least two elements in 340 stars with projected rotational velocities \vsini\ $<$ 30 km s$^{-1}$, where line blending does not severely affect the spectra. In addition, we analyzed the spectrum of Vesta to determine the solar reference abundances used in this study.

The Orion stars exhibit sub-solar [X/H] abundance ratios, in agreement with previous findings \citep{kos21}. We also find chemical homogeneity across the subgroups Orion A, B, $\lambda$ Ori, and Orion OB in all six elements analyzed. To provide context for the Orion abundances, we determined chemical abundances for a nearby sample of main-sequence stars, representing the solar neighborhood. While the median [X/H] values in Orion are broadly consistent with those in the solar neighborhood, a Kolmogorov–Smirnov (KS) test indicates that the two populations are statistically distinct.

Among the elements studied, three are $\alpha$-elements (Mg, Si, Ti), allowing us to compute [$\alpha$/Fe] ratios for both the young Orion stars and the nearby main-sequence sample. We find that Orion members have systematically lower median [$\alpha$/Fe] values compared to solar neighborhood stars, which may reflect the chemical evolution of the Galaxy.

Finally, we compared the median carbon abundance in Orion A with measurements from the Orion Nebula (M42) based on recombination lines \citep{mendez22}. We find good agreement when the logarithmic abundance in M42 is referenced to the solar value from \citet{l03}.

The atmospheric parameters and chemical abundances presented here represent a first step toward a homogeneous chemical characterization of young stars. Future work should aim to include stars with low to moderate infrared excesses, as well as to explore chemical abundances in fast rotators. This may be achieved through spectral deblending techniques and/or higher spectral resolution. These populations—stars with infrared excess and high rotation rates—represent a substantial fraction of the young stellar population in Orion and are essential for achieving a comprehensive understanding of the region's chemical evolution.


\section*{Acknowledgements}
We thank the referee for their helpful comments. R. L-V. and L.A. acknowledge support from Secretar\'ia de Ciencia, Humanidades, Tecnolog\'ia e Inovacci\'on (SECIHTI) through a postdoctoral fellowship within the program ``Estancias posdoctorales por M\'exico''. S.V. gratefully acknowledges the support provided by Fondecyt Regular n. 1220264 and by the ANID BASAL project FB210003. JEM-D acknowledges financial support from the UNAM/DGAPA/PAPIIT/IG101025 and UNAM/DGAPA/PAPIIT/IG104325 grants. C. R-Z., R. L-V., J.H., L.A. and J.S. acknowledge financial support from UNAM/DGAPA/PAPIIT/IG101723 and SECIHTI nd project CONAHCYT Ciencia Frontera 86372 (entitled Citlalcoatl).
Funding for the Sloan Digital Sky Survey IV has been provided by the Alfred P. Sloan Foundation, the U.S. Department of Energy Office of Science, and the Participating Institutions. SDSS-IV acknowledges support and resources from the Center for High-Performance Computing at the University of Utah. The SDSS website is www.sdss.org. 
SDSS-IV is managed by the Astrophysical Research Consortium for the Participating Institutions of the SDSS Collaboration including the Brazilian Participation Group, the Carnegie Institution for Science, Carnegie Mellon University, the Chilean Participation Group, the French Participation Group, Harvard-Smithsonian Center for Astrophysics, Instituto de Astrof\'isica de Canarias, The Johns Hopkins University, Kavli Institute for the Physics and Mathematics of the Universe (IPMU) / University of Tokyo, Lawrence Berkeley National Laboratory, Leibniz Institut f\"ur Astrophysik Potsdam (AIP), Max-Planck-Institut f\"ur Astronomie (MPIA Heidelberg), Max-Planck-Institut f\"ur Astrophysik (MPA Garching), Max-Planck-Institut f\"ur Extraterrestrische Physik (MPE), National Astronomical Observatories of China, New Mexico State University, New York University, University of Notre Dame, Observat\'ario Nacional / MCTI, The Ohio State University, Pennsylvania State University, Shanghai Astronomical Observatory, United Kingdom Participation Group, Universidad Nacional Aut\'onoma de M\'exico, University of Arizona, University of Colorado Boulder, University of Oxford, University of Portsmouth, University of Utah, University of Virginia, University of Washington, University of Wisconsin, Vanderbilt University, and Yale University.  
This research has made use of the SIMBAD database, operated at CDS, Strasbourg, France.
This work made use of the high-throughput computing (HTC) infrastructure and the optimisation of computing resources of the project Grid UNAM (\url{https://grid.unam.mx/}) at the Universidad Nacional Autónoma de México. Participating Institutions of the project Grid UNAM includes Dirección General de Cómputo y de Tecnologías de Información y Comunicación (DGTIC), Laboratorio de Modelos y Datos Científicos (LAMOD), Instituto de Ciencias de la Atmósfera y Cambio Climático (ICAyCC), Instituto de Astronomía (IA) and Instituto de Ciencias Nucleares (ICN). We also thank the members of technical and development staff of the project Grid UNAM for their continuous support.
\section*{Data Availability}
Table \ref{tab:res} is available in the online supplementary material of the paper.



\bibliographystyle{mnras}
\bibliography{biblio} 








\bsp	
\label{lastpage}
\end{document}

%% file: space.tex
\begin{table}
\caption{Parameter space of the MARCS synthetic grid used in this work.}
\label{tab:space}
\begin{tabular}{crl}
\hline
{Parameter} & {Range} & {Step} \\ 
\hline
\multirow{2}{*}{\teff} & \multirow{2}{*}{3000 --- 8000 K} & 100 K for \teff\ $\leq$ 4000 K \\ 
& & 250 K otherwise \\
\logg & 0.0 --- 5.5 dex & 0.5 dex \\
\meta & $-$2.50 --- 1.00 dex & 0.25 dex \\
{\alfa} & {$-$0.75 --- 1.00 dex} & 0.25 dex\\
\hline
\end{tabular}
\end{table}

%% file: resul.tex
\begin{table}
\caption{Descriptive content of the Table of the stellar parameters and chemical abundances determined in this work. The full table is available in the electronic version of the paper.}
\label{tab:res}
\begin{tabular}{ll}
\hline
{Column name} & {Description} \\        
\hline
OBJ  &  2MASS identification number\\
RA   &  Right Ascencion (J2000) \\
Dec  &  Declination (J2000) \\
t25  &  Effective temperature at percentile 25\\
t50  &  Effective temperature at percentile 50 \\
t75  &  Effective temperature at percentile  75\\
g25  &  Surface gravity at percentile 25 \\
g50  &  Surface gravity at percentile 50 \\
g75  &  Surface gravity at percentile 75\\
m25  &  Overall metallicity at percentile 25 \\
m50  &  Overall metallicity at percentile 50 \\
m75  &  Overall metallicity at percentile 75 \\
a25  &  alpha elements abundance at percentile  25\\
a50  &  alpha elements abundance at percentile 50 \\
a75  &  alpha elements abundance at percentile 75 \\
v25  &  Projected rotational velocity at percentile 25 \\
v50  &  Projected rotational velocity at percentile 50 \\
v75  &  Projected rotational velocity at percentile 75 \\

C\_H &  Mean [C/H] abundance \\
lowC &  lower error in carbon abundance \\
uppC &  Upper error in carbon abundance \\
lC   &  number of lines used in carbon abundance determination \\

Mg\_H&  Mean [Mg/H] abundance \\
lowMg&  lower error in magnesium abundance \\
uppMg&  Upper error in magnesium abundance \\
lMg  &  lines used in magnesium abundance determination \\

Si\_H&  Mean [Si/H] abundance \\
lowSi&  lower error in silicon abundance \\
uppSi&  Upper error in silicon abundance \\
lSi  &  lines used in silicon abundance determination \\

K\_H &  Mean [K/H] abundance \\
lowK &  lower error in potassium abundance \\
uppK &  Upper error in potassium abundance \\
lK   &  lines used in potassium abundance determination \\

Ti\_H&  Mean [Ti/H] abundance \\
lowTi&  lower error in titanium abundance \\
uppTi&  Upper error in titanium abundance \\
lTi  &  lines used in titanium abundance determination \\

Fe\_H&  Mean [Fe/H] abundance \\
lowFe&  lower error in iron abundance \\
uppFe&  Upper error in iron abundance \\
lFe  &  lines used in iron abundance determination \\
{\bf vmic} & microturbulence velocity \\
{\bf alfa\_Fe }&  Alpha-elements to iron abundance ratio \\
{\bf e\_alfa\_Fe}&  error in alfa\_Fe abundance \\ 
\hline
\end{tabular}
\end{table}

%% file: tab_lines.tex
\begin{table}
\caption{Basic information of the 19 atomic lines used in this study. The last column is the solar abundance determined using the APOGEE DR17 Vesta asteroid spectrum.} \label{tab:lines}
\begin{tabular}{cccccc}
\hline
{$\lambda$} & {element} & {$\chi$} & {$\log$ gf} & {Van der Wals} & A(X)$_{\rm Vesta}$\\
{(\AA)} &  & {(eV)} & {} & {} \\        
\hline
15784.536 & C  & 9.631 & -0.588  & -6.99 & 8.56$\pm$0.02\\
\hline
15740.716 & Mg & 5.931 & -0.323  & -6.84 & 7.51$\pm$0.06\\
15879.521 & Mg & 5.946 & -2.102  & -6.91 & 7.66$\pm$0.02\\ 
\hline
15884.454 & Si & 5.954 & -0.945  & -7.11 & 7.64$\pm$0.02\\
16060.009 & Si & 5.954 & -0.452  & -7.07 & 7.42$\pm$0.02\\
16094.787 & Si & 5.964 & -0.088  & -7.10 & 7.51$\pm$0.05\\
16215.670 & Si & 5.954 & -0.565  & -7.21 & 7.46$\pm$0.02\\ 
16241.833 & Si & 5.964 & -0.762  & -7.13 & 7.50$\pm$0.02\\
16828.159 & Si & 5.984 & -1.058  & -7.10 & 7.62$\pm$0.04\\
\hline
15163.067 & K & 2.670 &  0.632  & -6.82  & 4.93$\pm$0.02\\
15168.377 & K  & 2.670 &  0.441  & -6.82 & 5.06$\pm$0.02\\
\hline
16635.158 & Ti & 2.345 & -1.771  & -7.49 & 5.05$\pm$0.05\\
\hline
15207.526 & Fe & 5.385 &  0.067  & -7.21 & 7.65$\pm$0.03\\
15244.974 & Fe & 5.587 & -0.134  & -7.01 & 7.49$\pm$0.03\\
15723.586 & Fe & 5.621 & -0.011  & -7.05 & 7.51$\pm$0.03\\
15904.324 & Fe & 6.365 & -0.154  & -7.19 & 7.43$\pm$0.02\\ 
15920.642 & Fe & 6.258 & -0.011  & -7.01 & 7.45$\pm$0.02\\
16125.899 & Fe & 6.351 &  0.618  & -6.91 & 7.54$\pm$0.03\\
16316.320 & Fe & 6.280 &  0.857  & -6.90 & 7.49$\pm$0.03\\

\hline
\end{tabular}
\end{table}

%% file: tab_groups.tex
\begin{table*}
\caption{Median abundance values for each [X/H] and for each Orion group. We present the number of stars included in the computation,\#, followed by the median and median absolute deviation of our [X/H] estimations. \label{tab:groups}}
\begin{tabular}{lrccccccccccc}\hline
Group & \multicolumn{2}{c}{[C/H]} & \multicolumn{2}{c}{[Mg/H]} & \multicolumn{2}{c}{[Si/H]} & \multicolumn{2}{c}{[K/H]} & \multicolumn{2}{c}{[Ti/H]} & \multicolumn{2}{c}{[Fe/H]}\\ 
      & \# & $\tilde{x}\pm$MAD & \# & $\tilde{x}\pm$MAD & \# & $\tilde{x}\pm$MAD & \# & $\tilde{x}\pm$MAD & \# & $\tilde{x}\pm$MAD & \# & $\tilde{x}\pm$MAD\\
\hline
$\lambda$ Ori &  74 & -0.07$\pm$0.04 &  78 & -0.23$\pm$0.06 &  77 & $-$0.17$\pm$0.07 &  71 & -0.10$\pm$0.07 &  76 &-0.17$\pm$0.07&79 & -0.05$\pm$0.04 \\
Orion A       &  29 & -0.08$\pm$0.03 &  51 & -0.27$\pm$0.08 &  52 & $-$0.16$\pm$0.09 &  32 & -0.12$\pm$0.04 &  37 &-0.18$\pm$0.09&52 & -0.06$\pm$0.06 \\
Orion B       &  43 & -0.04$\pm$0.04 &  51 & -0.22$\pm$0.07 &  50 & $-$0.12$\pm$0.09 &  44 & -0.17$\pm$0.07 &  47 &-0.15$\pm$0.07&51 & -0.03$\pm$0.06 \\
Orion OB      & 139 & -0.06$\pm$0.03 & 152 & -0.23$\pm$0.04 & 154 & $-$0.14$\pm$0.05 & 139 & -0.07$\pm$0.06 & 138 &-0.12$\pm$0.06&155 & -0.04$\pm$0.04 \\
\hline
\end{tabular}
\end{table*}

%% file: ism_ab.tex
\begin{table*}
\caption{Median [X/H] values and MADs for Orion young stars and the main-sequence sample from \citet{lopezval24}, calculated in two different \teff\ bins. Columns 4 and 7 show the median KS test p-values obtained through a bootstrapping procedure comparing both populations. Only stars in the range $-0.5 \leq$ [Fe/H] $\leq +0.5$ were included in the comparisons.}
\label{tab:ism_comp}
\begin{tabular}{lcccccc}
\hline
&\multicolumn{3}{c}{\teff\ $\leq$ 4000}&\multicolumn{3}{c}{4000 $<$ \teff\ $\leq$ 5000}\\
$[{\rm X/H}]$ & Orion stars & main-sequence stars& $\mathcal{P}$(KS) & Orion stars & main-sequence stars& $\mathcal{P}$(KS)\\
\hline
C  & -0.05$\pm$0.04 (148) & -0.17$\pm$0.11 (327) & 2.4e-15 & -0.07$\pm$0.02 (128)& -0.04$\pm$0.09 (337) & 3.3e-06 \\
Mg & -0.22$\pm$0.06 (183) & -0.26$\pm$0.09 (441) & 1.6e-03& -0.24$\pm$0.04 (137)& -0.15$\pm$0.09 (401)&3.5e-12\\
Si & -0.18$\pm$0.08 (183) & -0.22$\pm$0.11 (449) & 7.3e-03& -0.12$\pm$0.04 (138) & -0.10$\pm$0.09 (411) & 1.5e-05 \\
K  & -0.13$\pm$0.08 (149) & -0.02$\pm$0.08 (348) & 2.1e-09&-0.08$\pm$0.04 (128) & -0.06$\pm$0.08 (404)& 1.0e-03\\
Ti & -0.16$\pm$0.08 (163) & -0.23$\pm$0.10 (418) & 1.2e-05& -0.13$\pm$0.07 (127) & -0.05$\pm$0.10 (400)& 1.5e-06\\
Fe & -0.04$\pm$0.05 (188) & -0.18$\pm$0.13 (647) & 1.1e-17& -0.04$\pm$0.04 (139)& -0.05$\pm$0.10 (411)& 1.6e-04\\
\hline
\end{tabular}
\end{table*}